\documentclass[12pt]{article}
\usepackage{graphicx}
\usepackage{amsmath}
\usepackage{amsfonts}
\usepackage{eucal}
\usepackage{mathbbol}

\def\pbnr{}
\def\speaker{Timothy Burns}
\def\title{Angular momentum coefficients for meson strong decay and the unquenched quark model}
\def\affiliation{Department of Mathematical Sciences, Durham University, Durham DH1 3LE, United Kingdom}
\def\support{{\tt t.burns@oxon.org}}
\newcommand\pubnumber{\pbnr}
\newcommand\pubdate{\today}
\def\Title#1{\begin{center} {\Large #1 } \end{center}}
\def\Author#1{\begin{center}{ \sc #1} \end{center}}

\newcommand{\OnBehalf}[1]{\sbox0{#1}\ifdim\wd0=0pt
        {}% if #1 is empty
	\else
	{\\on behalf of #1}% if #1 is not empty
	\fi}
\newcommand{\SupportedBy}[1]{\sbox0{#1}\ifdim\wd0=0pt
        {}% if #1 is empty
	\else
	{\footnote{#1}}% if #1 is not empty
	\fi}
\def\Address#1{\begin{center}{ \it #1} \end{center}}

\newcommand\pubblock{\includegraphics[width=5cm]{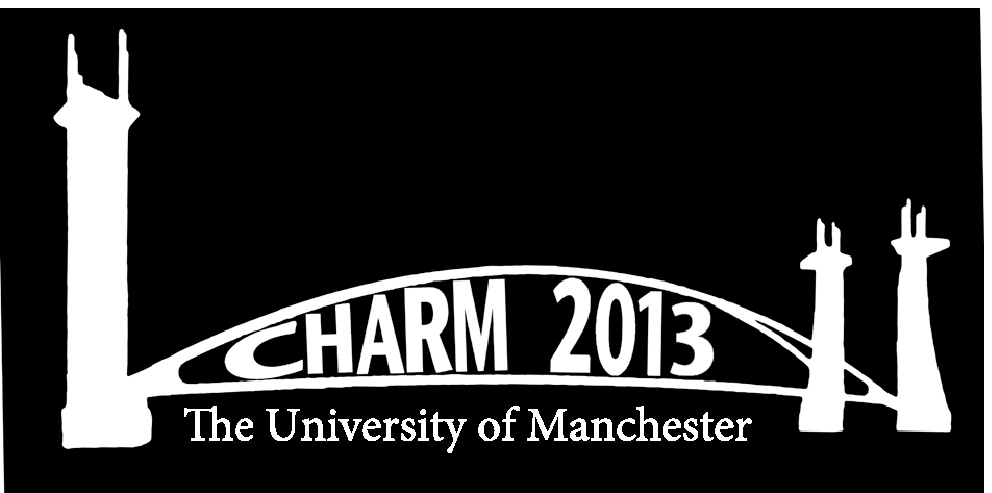}\hfill{\begin{tabular}{l} \pubnumber\\
         \pubdate  \end{tabular}}}
\newenvironment{Abstract}{\begin{quotation}  }{\end{quotation}}
\newenvironment{Presented}{\begin{quotation} \begin{center} 
             PRESENTED AT\end{center}\bigskip 
      \begin{center}\begin{large}}{\end{large}\end{center} \end{quotation}}

\def\venue{The 6$^{th}$ International Workshop on Charm Physics\\
(CHARM 2013)\\
Manchester, UK,  31 August -- 4 September, 2013}
\def\beq{\begin{equation}}
\def\eeq#1{\label{#1}\end{equation}}
\def\eeqn{\end{equation}}

\def\beqa{\begin{eqnarray}}
\def\eeqa#1{\label{#1}\end{eqnarray}}
\def\eeqan{\end{eqnarray}}

\let\bar=\overbar

\def\Dslash{\not{\hbox{\kern-4pt $D$}}}
\def\dslash{\not{\hbox{\kern-2pt $\del$}}}

\def\ee{e^+e^-}

\def\msb{{\bar{\ssstyle M \kern -1pt S}}}

\def\qq{q\overline q}
\def\q{\overline q}
\def\Q{\overline Q}

\def\>{\big>}
\def\<{\big<}
\def\|{\big|}
\newcommand{\ve}[1]{\mathbf{#1}} 	%roman vector
\newcommand{\vh}[1]{\mathbf{\hat{#1}}}	% v-hat
\renewcommand{\u}[1]{\rm{#1}}
\def\cn#1#2#3{^{#1}{\u{#2}}_{#3}}     %normal chemical notation
\def\an{\cn} %atomic notation
\newcommand{\uS}{\rm{S}}
\newcommand{\uP}{\rm{P}}
\newcommand{\uD}{\rm{D}}
\newcommand{\uF}{\rm{F}}

\def\be{\begin{equation}} 
\def\ee{\end{equation}}  
\def\bea{\begin{eqnarray}}  
\def\eea{\end{eqnarray}} 
\def\smallwigner#1{\!\left\{\!\begin{smallmatrix}#1\end{smallmatrix}\!\right\}}

 \def\spato{\mathbf{ O}}
 
 \def\spino{\boldsymbol{\chi}}

 \def\p{_{\phantom{0}}}
 \def\ot{\times}

 \def\qnset#1{\!\bigg[\!\begin{smallmatrix}#1\end{smallmatrix}\!\bigg]}

\def\MeV{\textrm{ MeV}}

\def\epem{e^+e^-}
\def\hyperfine#1{\Delta M_{#1\uS}}

\def\epemwidth#1{\Gamma_{{\epem}\to{#1^3 \rm S_1}}}
\begin{document}
\begin{titlepage}
\pubblock

\vfill
\Title{\title}
\vfill
\Author{\speaker\SupportedBy{\support}}
\Address{\affiliation}
\vfill
\begin{Abstract}

Most models for the strong decay of mesons, as well as unquenched quark models which incorporate the effect of coupling to meson-meson channels, assume that the coupling is driven by the creation of a $\qq$ pair in spin triplet. Their matrix elements can be factorised into a sum over (model-dependent) spatial matrix elements multiplied by (model-independent) coefficients. A general expression for these $\xi$ coefficients is obtained, and their properties are studied. Numerical tables of the coefficients can be used as a starting point for future calculations. The coefficients lead to model-independent predictions for decay amplitudes and widths, and can explain how mass shifts in the unquenched quark model do not spoil successful predictions of the ordinary (quenched) quark model. This article is based on work which will be presented in a forthcoming paper \cite{tjbinpreparation}.

\end{Abstract}
\vfill
\begin{Presented}
\venue
\end{Presented}
\vfill
\end{titlepage}
\def\thefootnote{\fnsymbol{footnote}}
\setcounter{footnote}{0}
%

%%%%%%%%%%%%%%%%%%%%%%%%%%%%%%%%%%%%%%%%%%%%%%%%%%%%%%%%%%%%%%%%%%%%%%%%%%%
%  WHAT FOLLOWS IS YOUR TEXT
%%%%%%%%%%%%%%%%%%%%%%%%%%%%%%%%%%%%%%%%%%%%%%%%%%%%%%%%%%%%%%%%%%%%%%%%%%%

\section{Decay Models}

Models for the transition $(Q\Q)\to (Q\q)(q\Q)$ typically assume that the $q\q$ pair is created in spin-one, although they differ in the treatment of the spatial degrees of freedom. One approach argues that the created pair must have the $0^{++}$ quantum numbers of the vacuum, and so is created in a  $\an 3P0$ state (see, for example, refs \cite{Micu:1968mk,LeYaouanc:1972ae}). The amplitude to produce a $q$ and $\q$ with projections $s$ and $\overline s$ of quark spin, and momenta $\ve k$ and $\ve{\overline k}$, factorises into a scalar product of vectors associated with the spin and spatial parts,
\bea
\<s \ve k , \overline s\ve{\overline k}\|T\|0\>
&=&\spino_{s\overline s}\cdot\spato(\ve k,\ve{\overline k}),
\label{momop}\\
\spato(\ve k,\ve{\overline k})&=&(\ve {\overline k}- \ve{k})\delta^3(\ve {\overline k}+\ve{k}),
\eea
where $\spino_{s\overline s}$ is the wavefunction of a spin triplet $\qq$, pair and $\spato$ contains the spatial dependence. 

The flux tube model \cite{Kokoski:1985is} also involves a $\qq$ pair in spin triplet and the factorisation is the same. The only difference is the spatial part, which is modified by the overlap $\gamma(\ve x)$ of the flux tubes of the mesons. For a discrete flux tube which breaks between $\ve x$ and $\ve x +a\vh n$, 
\be
\spato(\ve k, \ve{\overline k})
\approx
\frac{\sqrt 2}{(2\pi)^3}
\vh n
(1-ia\vh n\cdot\ve{\overline k})
\int d^3\ve x
\gamma(\ve x)
e^{-i(\ve k+\ve{\overline k})\cdot \ve x},
\ee 
and the two terms create a $\qq$ pair with $\an 3S1$ and $\an 3P0$ quantum numbers respectively. 

The Cornell model \cite{Eichten:1975ag,Eichten:1978tg}, and the dominant terms in a more general microscopic model \cite{Ackleh:1996yt}, also have the same factorisation with the $q\q$ created in spin triplet. In these cases, scattering off an initial quark involves a momentum transfer $\ve K'-\ve K$ but crucially leaves its spin projection unchanged; the spatial part involves the Fourier transform $\widetilde V$ of the interaction potential and the light quark mass $m$:
\be
% \<S\ve K',s\ve k,\overline s\ve{\overline k} \|T \| S\ve K\>
% &=&
% \spino_{s\overline s}
% \cdot
% \spato(\ve K, \ve K',\ve k,\ve{\overline k})
% \label{microdot}
% \\
\spato(\ve K, \ve K',\ve k,\ve{\overline k})
=
\frac{\sqrt 2}{2m}
(\ve{\overline k}\pm\ve k)
\delta^3(\ve K-\ve K'-\ve k-\ve{\overline k})
\widetilde V(\ve k+\ve{\overline k}).
\label{sKs}
\ee

In the pion emission model \cite{Becchi:1966zz,Mitra:1967zz} (and pseudoscalar-meson emission models more generally) one ignores the quark sub-structure of one of the final states entirely. Nevertheless if the transition is interpreted in terms of pair creation, one arrives at an equivalent physical picture to the models described above: a $q\q$ pair is created in spin triplet, and the initial quark spins act as spectators. 

There are many different variants on each of these ``non-flip, triplet'' models, but they differ only in the spatial degrees of freedom. Because they have the same spin structure, the angular momentum dependence of their matrix elements is the same. 

\section{Angular Momentum Coefficients}

The idea is to obtain a general solution for the matrix element in the above models which isolates the common angular momentum dependence. Consider an arbitrary transition
\be
nSLJ\to n_1S_1L_1J_1+n_2S_2L_2J_2,
\ee
where $n$ is the radial quantum number, and $S$, $L$ and $J$ are the spin, orbital and total angular momenta. For final states coupled to angular momentum $j$ and in a relative partial wave $l$ the expression is
\be
M_{jl}
\qnset{n\p S\p L\p J\p\\n_1S_1L_1J_1\\n_2S_2L_2J_2}_\pm
=
\sum_{L'l'}
\xi_{jl}^{L'l'}
\qnset
{S\p L\p J\p\\
S_1L_1J_1\\
S_2L_2J_2}_\pm
A_{l}^{L'l'}
\qnset
{n\p L\p\\
n_1L_1\\
n_2L_2}_\pm.
\label{eq:defining}
\ee
The summation variables $L'$ and $l'$ are quantum numbers associated with the orbital couplings, and are not relevant in most cases of phenomenological interest. The $\pm$ labels refer to two possible topologies distinguished by the arrangement of the initial and created (anti)quarks in the final state mesons.

The term $A$ is the reduced matrix element of the spatial part $\spato_\pm$ of the operator, 
\be
A_{l}^{L'l'}
\qnset
{n\p L\p\\
n_1L_1\\
n_2L_2}_\pm
=\frac{1}{|L|}\<((n_1L_1\ot n_2L_2)_{L'}\ot l)_{l'}\|\|\spato_\pm\|\|nL\>,
\label{spatialme}
\ee
where $\|L\|=\sqrt{2L+1}$. It is a function of the decay momenta obtained by integrating over the spatial wavefunctions of the mesons, and contains all of the model-dependence. 

Conversely the angular momentum coefficients $\xi$ are common to all non-flip, triplet models, and depend only on the angular momenta of the mesons involved. A general expression for $\xi$ can be obtained in terms of Wigner-Racah coefficients, which arise due to the recoupling and subsequent factorisation of the spin and orbital degrees of freedom,
\begin{multline}
\xi_{jl}^{L'l'}
\qnset
{S\p L\p J\p\\
S_1L_1J_1\\
S_2L_2J_2}_\pm
=
\sum_{S'}(-)^{S+S'+l+l'+L'} 
|J_1J_2S'L'jl'L|
\smallwigner{S_1&L_1&J_1\\S_2&L_2&J_2\\S'&L'&j}
\smallwigner{S'&L'&j\\l&J&l'}
\smallwigner{S &L &J \\l'&S'&1}
\\
\times
\<(S_1\ot S_2)_{S'}\|\|\spino_\pm\|\|S\>,
\label{eq:xi}
\end{multline}
where $|J_1\ldots L|=|J_1|\ldots|L|$, and the spin matrix element is
\be
\<(S_1\ot S_2)_{S'}\|\|\spino_+\|\|S\>=(-)^{S_2+1}|S_1S_2S'S1|\smallwigner{1/2&1/2&S_1\\1/2&1/2&S_2\\S&1&S'}.
\ee

The Wigner-Racah approach outlined above has appeared elsewhere in the literature, within the context of specific models \cite{BonnazSilvestre-Brac99discussion,Burns:2006rz,Burns:2007hk}. The discussion in the previous section implies that the angular momentum algebra applies more generally to all non-flip, triplet models.

%  \begin{center}\vspace{2cm}
% % \begin{figure}
% 
% \includegraphics[width=0.85\linewidth]{topo}
%   \rput(-25,5.5){Topology $(+)$}
%   \rput(-34,4){$SLJ$}
%   \rput(-18,4){$S_1L_1J_1$}
%   \rput(-18,1){$S_2L_2J_2$}
%   \rput(-6,5.5){Topology $(-)$}
%   \rput(-14.5,4){$SLJ$}
%   \rput(1.2,4){$S_1L_1J_1$}
%   \rput(1.2,1){$S_2L_2J_2$}
% 
% % \captionof{figure}{\color{Green} Figure caption}
% % \end{figure}
%  \end{center}\vspace{1cm}

\section{Some basic properties}

The $\xi$ coefficients have the following symmetry properties under the interchange of the topologies, and of the quantum numbers of mesons 1 and 2:
\bea
\xi_{jl}^{L'l'}
\qnset
{S\p L\p J\p\\
S_1L_1J_1\\
S_2L_2J_2}_\pm
&=&
(-)^{S+S_1+S_2+1}
\xi_{jl}^{L'l'}
\qnset
{S\p L\p J\p\\
S_1L_1J_1\\
S_2L_2J_2}_\mp
\label{xisymm[1]}
\\
&=&
(-)^{S_1+L_1+J_1+S_2+L_2+J_2+S+L'+j+1}
\xi_{jl}^{L'l'}
\qnset
{S\p L\p J\p\\
S_2L_2J_2\\
S_1L_1J_1\\}_\pm.
\eea
The first of these can be shown to lead to the conservation of $G$-parity. The second leads to a new selection rule for final states with the same spin, orbital and total angular momenta.

The coefficients satisfy an  orthogonality relation which underpins the theorems of refs \cite{Barnes:2007xu,Close:2009ii} for mass shifts and mixing amplitudes in the unquenched quark model, 
\be
\sum_{\substack{S_1S_2\\J_1J_2j}}
\xi_{jl}^{\widehat L' \widehat l'}
\qnset
{\widehat S\p\widehat L\p J\p\\
S_1L_1J_1\\
S_2L_2J_2}_\pm
\xi_{jl}^{L'l'}
\qnset
{S\p L\p J\p\\
S_1L_1J_1\\
S_2L_2J_2}_\pm
=\delta_{\widehat S S}\delta_{\widehat L L}\delta_{\widehat L' L'}\delta_{\widehat l' l'}.
\label{orthogrelation}
\ee

\section{Strong decay}

The $\xi$ coefficients lead to model-independent predictions common to all non-flip, triplet models. A few examples are given here, concentrating on the decays of charmonia and charmed-mesons to $\an 1S0$ pseudoscalar ($P$) and $\an 3S1$ vector ($V$) mesons. Many more examples will be presented in a future paper \cite{tjbinpreparation}, where light meson decays are also considered.

Zeroes in the $\xi$s are selection rules. There is a new ``spin triplet'' selection rule, for example, related to the well-known spin singlet selection rule. The rule forbids the transition $\an 3P1\to VV$ in S-wave, but not D-wave; this could be tested in the $D^*\overline D^*$ decays of the $\chi_{c1}(3\uP)$, which has yet to be discovered.

Ratios of amplitudes with the same $l$ but with different $j$ are exact predictions of non-flip, triplet models, and these can be measured experimentally. For example $\an3P2\to VV$ in D-wave has $M_{2\uD}/M_{0\uD}=-\sqrt{7}$. ($M_{1\uD}=0$ due to the spin vector selection rule.) For charmonia this could be tested in the $D^*\overline D^*$ decays of the $\chi_{c2}(3\uP)$.

Matrix elements for mesons with the same $n$ and $L$ but different $S$ and $J$ involve the same spatial matrix element $A$, which leads to relations among their amplitudes and widths, valid in the limit of identical radial wavefunctions and equal decay momenta. For $VP$ decays, for example, the partial widths $\Gamma$ and F-to-P ratios satisfy the following relations:	
\be
3\Gamma(\an 3D1)+2\Gamma(\an 1D2)
=
3\Gamma(\an 3D2)\textrm{\qquad and}\qquad
\frac{M_{\uF}(\an 3D2)}{M_{\uP}(\an 3D2)}
\frac{M_{\uP}(\an 1D2)}{M_{\uF}(\an 1D2)}
=
-\frac{2}{3}.
\ee
These could eventually be tested in the $D^*\overline D$ decays of $2\uD$ charmonia, for example, where the equal momenta approximation is a good one. (Model predictions agree that the states are almost degenerate.) A similar relation for the decay of the $\pi_1$ hybrid meson is consistent with lattice QCD \cite{Burns:2006wz,Burns:2007hk}. In ref. \cite{Burns:2007dc} relations of a similar nature were employed in a model for exclusive decays of charmonia to light mesons.

Other ratios can be used to extract mixing angles, such as for heavy-light states of mixed spin. For example if $A$ and $B$ are orthogonal mixtures of $\an 3P1$ and $\an 1P1$, their mixing angle $\phi$ can be extracted from the ratio of D-to-S ratios in their $PV$ decays,
\be
\frac{M_{\uD}(A)}{M_{\uS}(A)}
\frac{M_{\uS}(B)}{M_{\uD}(B)}
=
-\tan^2(\phi-\phi_P),
\ee
where $\phi_{\uP}=35.3^\circ$ is the heavy quark mixing angle. (The zero at $\phi=\phi_{\uP}$ is due to the conservation of light quark spin.) These relations could be applied to the $D^*\pi$ decays of the 1P states $D(2420)$ and $D(2430)$ for which, owing to their near degeneracy, the equal momenta approximation is a good one. For the heavier $D(2\uP)$ mesons for which the $D^*\rho$ may be open, the ratios of D-wave $VV$ amplitudes are given by the exact expressions
\be
\frac{M_{2\uD}(A)}{M_{1\uD}(A)}=\cot \phi\sqrt{\frac{3}{2}}
\textrm{\qquad and}\qquad
\frac{M_{2\uD}(B)}{M_{1\uD}(B)}=-\tan\phi\sqrt{\frac{3}{2}}.
\ee
The $\xi$s could also be used to extract mixing angles for states of mixed orbital angular momenta; e.g. if $A$ and $B$ are mixtures of $\an 3S1$ and $\an 3D1$ with mixing angle $\phi$, the ratio of D-to-S ratios of ${\an 3P0}{\an 3S1}$ is
\be
\frac{M_{\uD}(A)}{M_{\uS}(A)}
\frac{M_{\uS}(B)}{M_{\uD}(B)}
=
-\tan^2\phi.
\ee

\section{The unquenched quark model}

The ``unquenched'' quark model aims to explain the puzzling properties of new mesons by accounting for their coupling to meson-meson thresholds. The challenge is that the model must also retain the empirically successful features of the ordinary (quenched) quark model. For example the prediction for zero hyperfine splitting of P-wave mesons 
\be
\frac{1}{9}\left(M_{\an 3P0}+3M_{\an 3P1}+5M_{\an 3P2}\right)-M_{\an 1\uP1}=0,\label{zerohyperfine}
\ee
is confirmed in experiment for 1P charmonia, and 1P and 2P bottomonia, e.g.
\be
\overline M_{\chi_c(1\uP)}-M_{h_c(1\uP)} =+0.02\pm 0.19\pm 0.13 \MeV \cite{Dobbs:2008ec}.\\
\ee
Coupling to meson-meson thresholds induces mass shifts which threaten to spoil this nice, model-independent prediction. Happily this is not the case: despite large and different mass shifts for each of the $\an 1P1$, $\an 3P0$, $\an 3P1$ and $\an 3P2$ states, their net contribution to the hyperfine splitting conspires to be very small. This mechanism is  common to non-flip, triplet models and can be understood in terms of the $\xi$s \cite{Burns:2011fu,Burns:2011jv}. The mass shifts of 1P charmonia in ref. \cite{Li:2009ad} are a typical example of the effect; their contribution (MeV) to the hyperfine splitting is:
\be
\frac{1}{9}(131+3\times152+5\times175)-162=0.4.
\ee
The relation equivalent to (\ref{zerohyperfine}) for D-wave mesons is also protected by the same mechanism. Using that relation one can predict the mass of the missing $\an 1D2$ bottomonium, which turns out to be consistent with the prediction of a string model \cite{Burns:2010qq}.

A second example is the relation in the ordinary (quenched) quark model between the hyperfine splittings of S-wave mesons and the $\epem$ widths of the $\an 3S1$ states, 
\be
\frac{\hyperfine{2}}{\hyperfine{1}}=\frac{\epemwidth{2}}{\epemwidth{1}},
\ee
which is consistent with data for charmonia and, following the recent discovery of the $\eta_b(2\uS)$ \cite{Mizuk:2012pb}, bottomonia. Coupling to meson-meson channels modifies both sides of the above equation and threatens to spoil the prediction.  The $\epem$ widths are suppressed by the probability $P$ that the physical states are $(Q\Q)$, rather than $(Q\q)(q\Q)$. Remarkably, the different mass shifts of the  $\an 1S0$ and $\an 3S1$ states lead to a suppression of the hyperfine splittings by the same factor $P$, so that the relation holds. The mechanism can also be explained by means of the  $\xi$ coefficients \cite{Burns:2012pc}.

\section{Conclusions}

Most approaches to meson strong decay and the unquenched quark model share a common angular momentum dependence, which can be parametrised very conveniently by $\xi$ coefficients. This article has provided a very short introduction to the properties of these coefficients and their applications. In a forthcoming paper \cite{tjbinpreparation} the $\xi$ coefficients are discussed in more detail, and tables of their values are presented as a tool for future calculations.

\bibliography{tjb}

\end{document}